\documentclass[12pt]{article}
\usepackage{epsfig}

\def\beq{\begin{equation}}
\def\eeq#1{\label{#1}\end{equation}}
\def\eeqn{\end{equation}}

\def\beqa{\begin{eqnarray}}
\def\eeqa#1{\label{#1}\end{eqnarray}}
\def\eeqan{\end{eqnarray}}

\let\bar=\overbar

\def\Dslash{\not{\hbox{\kern-4pt $D$}}}
\def\dslash{\not{\hbox{\kern-2pt $\del$}}}

\def\msb{{\bar{\ssstyle M \kern -1pt S}}}

\usepackage{fancyhdr,graphicx}
\def\Title#1{\begin{center} {\Large {\bf #1} } \end{center}}

\begin{document}

\Title{Vortex Dynamics in Color-Superconducting Quark stars:
The Re-heating of Magnetars}

\bigskip\bigskip

%+\addcontentsline{toc}{chapter}{{\it I. Med}}
%+\label{MedInseneStart}

\begin{raggedright}

{\it Brian Niebergal$^{1,2}$, Rachid Ouyed$^{1}$, Rodrigo
  Negreiros$^{2}$, Fridolin Weber$^{2}$ \\
$^1$Department of Physics and Astronomy, University of
  Calgary, 2500 University Drive NW, Calgary, Alberta, T2N 1N4, Canada\\
  $^2$Department of Physics, San Diego State University, 5500
  Campanile Drive, San Diego, California 92182, USA \\
{\tt Email: bnieber@phas.ucalgary.ca}}
\bigskip\bigskip
\end{raggedright}

\begin{abstract}
 Compact stars  made of quark matter rather than confined
  hadronic matter,  are expected to form a color
  superconductor. This superconductor ought to be threaded with
  rotational vortex lines within which  the star's interior magnetic field
   is confined.  The vortices (and thus magnetic flux) would be expelled from the star during
  stellar spin-down, leading to magnetic reconnection at
   the surface of the star and the prolific production of thermal
  energy.  In this Letter, we show that this energy release can
  re-heat quark stars to exceptionally high temperatures, such as
  observed for Soft Gamma Repeaters (SGRs), Anomalous X-Ray pulsars
  (AXPs), and X-ray dim isolated neutron stars (XDINs). Moreover, our
  numerical investigations of the temperature evolution, spin-down
  rate, and magnetic field behavior of such superconducting quark
  stars suggest that SGRs, AXPs, and XDINs may be linked ancestrally.
  Finally, we discuss the possibility of a time delay before the star enters the
  color superconducting phase, which can be used to estimate the density
  at which quarks deconfine.  We find this density to be five times 
  that of nuclear saturation.
\end{abstract}

The interiors of compact stars provide a naturally-occuring
environment that can be used for studying the properties of
ultra-dense baryonic matter. A common means of probing this
environment is through direct observations of thermal emission from
the surface of compact stars.  By comparing these observations with
theoretical models one can retrieve key information about the physical
processes occuring in matter compressed to ultra-high nuclear
densities
\cite{1996csnp.book.....G,1999Weber..book,2004ApJS..155..623P,%
  2004ARA&A..42..169Y,Weber:2004kj,2006NuPhA.777..497P}.

Most attempts to model thermal emission make use of what is called the
minimal cooling scenario, which involves cooling through the minimum
set of particle processes that are necessary to explain the thermal
evolution of the majority of compact stars.  Appropriately, it is used
as a benchmark for observations of cooling neutron stars.  However,
there are a number of compact stars (see Table \ref{tab:data})
possessing thermal emissions significantly out of agreement with the
minimal cooling scenario, indicating that other processes may be
occuring and need consideration. Some specific classes of compact
stars that disagree with minimal cooling, and are already distinct for
some of their other features, are Soft Gamma-ray Repeaters (SGRs), Anomalous X-ray
Pulsars (AXPs), and X-ray Dim Isolated Neutron stars (XDINs).  It is
generally accepted that SGRs and AXPs are the same type of objects,
and it has been speculated before that XDINs are also related
\cite{2000PASP..112..297T}.  Observations indicate that SGRs and AXPs
are very hot objects.  Heating by magnetic field decay in the crust
has been suggested by Pons et al.\ \cite{2007PhRvL..98g1101P} as a
possible explanation and, at certain times during the star's
evolution, it has been shown to possibly be the dominant source of
heating depending on the values assumed for magnetic field decay
timescales \cite{2008ApJ...673L.167A}.  However, even with the most
liberal decay timescale parameters, crustal magnetic field decay
does not produce enough heat to explain SGR and AXP observations.
In fact, there is no physical model that can explain the unusually 
high temperatures of the objects in Table \ref{tab:data}.  Phenomenological
studies have been performed \cite{2000ApJ...529L..29C}, 
which parametrize the magnetic field decay to fit observations, 
and conclude that some efficient mechanism of magnetic flux expulsion
from the star's core is required.  Other studies proceed by introducing an artifical
heat source in an internal layer \cite{2009MNRAS.395.2257K}.

In this Letter, we expand on the idea that magnetic field decay from
compact stars causes a re-heating of such objects. As stated above,
the net effect of this mechanism on the temperature of
ordinary neutron stars is much too weak to accommodate the
temperatures observed for SGRs, AXPs, and XDINs. The situation changes
dramatically if one assumes that these objects are made of
superconducting quark matter rather than confined hadronic matter. 
As demonstrated in our previous work \cite{2007A&A...476L...5N} and below, 
magnetic flux expulsion from the cores of 
such stars provides a very efficient and
robust mechanism that can re-heat compact stars to the temperature
regime observed for AXPs, SGRs, and XDINs. 

\begin{table}[t!]
\caption{\label{tab:data} Compact stars with reported 
thermal emissions significantly higher than can be predicted by the 
minimal cooling scenario, specifically SGRs, AXPs, and XDINs 
\cite{2008ApJ...673L.167A}. The listed ages are the spin-down 
ages modified by the vortex expulsion process ($P/3\dot{P}$), 
instead of the usually assumed spin-down age ($P/2\dot{P}$).}
\begin{tabular}{ccc}
Name & $T \times 10^6$     & Age    \\
     & (K)                 & ($10^3$~years)  \\
\hline
SGR 1806-20 &  $7.56^{+0.8}_{-0.7}$ & $0.15$ \\
1E 1048.1-5937 & $7.22^{+0.13}_{-0.07}$ &$2.5$ \\
CXO J164710.2-455216 & $7.07$ & $0.5$ \\
SGR 0526-66 &  $6.16^{+0.07}_{-0.07}$ & $1.3$ \\
1RXS J170849.0-400910 & $5.3^{+0.98}_{-1.23}$ & $6.0$ \\
1E 1841-045 & $5.14^{+0.02}_{-0.02}$ & $3.0$ \\
SGR 1900+14  & $5.06^{+0.93}_{-0.06}$ & $0.73$ \\
1E2259+586$^\dagger$  &$4.78^{+0.34}_{-0.89}$ & $153$ \\
4U0142+615$^\dagger$  &$4.59^{+0.92}_{-0.40}$ & $47$ \\
CXOU J010043.1-721134 & $4.44^{+0.02}_{-0.02}$ & $4.5$\\
XTE J1810-197 & $7.92^{+0.22}_{-5.83}$ & $11.3$ \\
RX J0720.4-3125 & $1.05^{+0.06}_{-0.06}$ & $1266$ \\
RBS 1223 & $1.00^{+0.0}_{-0.0}$ & $974$ \\
\end{tabular}
\footnotetext{$^\dagger$1E2259+586 and 4U0141+615 are not considered here. As argued
in \cite{2007A&A...475...63O} these harbour an accreting ring that
explains their extreme luminosities.}
\end{table}

The feature that quark
matter forms a color superconductor %exhibiting the Meissner effect
is critical in our study. The condensation pattern that is considered
here is the color-flavor-locked (CFL)
phase \cite{2000hep.ph...11333R,2001ARNPS..51..131A,%
  2008RvMP...80.1455A,2001PhRvL..86.3492R} where quarks of all colors
and flavors pair together to form Cooper pairs.  Henceforth we refer
to such stars as color-flavor-locked quark stars 
(CFLQS) \footnote{Other condensation patterns for quark
matter fail to produce enough re-heating
\cite{2002PhRvD..65h5009F,2003NuPhA.718..697I,2006PhRvD..73g4009B} 
while pairing and superconductivity in neutron matter is also insufficient
\cite{2008AIPC..983..379Y}. }.  
Because of stellar rotation, CFLQSs develop rotationally-induced vortices.
The total number of vortices at any given time is $N_v =
\kappa\Omega$, with $\kappa$ being the vortex circulation and $\Omega$ the 
star's spin angular frequency.  Hence, as such a star spins-down
the number of vortices decrease.  They do so by being forced 
radially outward \cite{2004A&A...420.1025O} and upon reaching 
the surface are expelled from the star. The magnetic field, 
which exists only inside the vortices, is then also expelled and 
the subsequent magnetic reconnection leads to the production of X-rays 
\cite{2006ApJ...653..558O}. 
Since the star's spin-down rate is proportional to the magnetic field 
strength, when vortices are expelled, along with the magnetic 
flux contained within them, the spin-down rate also decreases.
Thus, the spin-down rate and magnetic field strength of a CFLQS 
become coupled.

The X-ray luminosity from this vortex expulsion process is given
by \cite{2006ApJ...646L..17N}
\begin{equation}\label{eq:lx}
  L_{\rm X} \simeq 2.01 \times 10^{34} \ {\rm erg\ s}^{-1} \ 
  \eta_{\rm X, 0.1} \ \dot{P}_{-11}^2 \ ,
\end{equation}
where $\eta_{\rm X, 0.1}$ is the re-connection efficiency parameter in units of
$0.1$ and $\dot{P}$ the spin-down rate in units of $10^{-11}$ s
s$^{-1}$. An estimate for the latter as well as the magnetic field evolution, 
both derived from vortex expulsion, is
\begin{equation}\label{eq:pdot}
  \dot{P} = \frac{P_0}{3 \tau} \left( 1 + \frac{t}{\tau} \right)^{-2/3} \ , 
  B = B_0\left(1+\frac{t}{\tau}\right)^{-1/6} \ ,
\end{equation}
where $P_0$ denotes the rotational period of the star at birth, $B_0$ the
magnetic field at birth, and $\tau$ is a relaxation time given by
\begin{equation}
  \tau = 5\times10^4 \left(\frac{10^{14} {\rm G}}{B_0}\right)^2 \left(\frac{P_0}{5 {\rm s}}
  \right)^2 \left(\frac{M}{M_{\odot}}\right)\left(\frac{10 {\rm km}}{R}\right)^{4} 
  \rm{yrs} \ .
  \label{tau}
\end{equation}
One can see from equations (\ref{eq:lx} \& \ref{eq:pdot}) 
that the X-ray emission decreases
as the star spins down. This X-ray emission is produced on the surface
of the CFLQS, which alters its thermal evolution. 
To study this numerically, the general relativistic equations of energy
balance and thermal energy transport need to be solved. These equations
are given by ($G = c = 1$)
\begin{eqnarray}
  \frac{ \partial (l e^{2\phi})}{\partial m}& = 
  &-\frac{1}{\rho \sqrt{1 - 2m/r}} \left( \epsilon_\nu 
    e^{2\phi} + c_v \frac{\partial (T e^\phi) }{\partial t} \right) \, , 
  \label{coeq1}  \\
  \frac{\partial (T e^\phi)}{\partial m} &=& - 
  \frac{(l e^{\phi})}{16 \pi^2 r^4 \kappa \rho \sqrt{1 - 2m/r}} 
  \label{coeq2} 
  \, ,
\end{eqnarray}
respectively \cite{1999Weber..book}. Here, $r$ is the distance from
the center of the star, $m(r)$ is the mass, $\rho(r)$ is the energy
density, $T(r,t)$ is the temperature, $l(r,t)$ is the luminosity,
$\phi(r)$ is the gravitational potential, $\epsilon_\nu(r,T)$ is the
neutrino emissivity, $c_v(r,T)$ is the specific heat, and
$\kappa(r,T)$ is the thermal conductivity \cite{1999Weber..book}. The
boundary conditions of (\ref{coeq1}) and (\ref{coeq2}) are determined
by the luminosity at the stellar center and at the surface. The
luminosity vanishes at the stellar center since there is no heat flux
there.  At the surface, the luminosity is defined by the relationship
between the mantle temperature and the temperature outside of the star
\cite{2000ApJ...533..406B}. 

Heating/cooling mechanisms used in our work, 
other than those included in the minimal cooling scenario, 
are the quark direct Urca process, to emissivity of which is on the order of $10^{26}
~\rm{erg}~ \rm{s}^{-1}~ \rm{cm}^{-3}$, and the quark modified Urca
and Bremsstrahlung, which are of order $10^{19}~ \rm{erg}~ \rm{s}^{-1}~
\rm{cm}^{-3}$. Due to quark pairing in the CFL state, however, the
direct Urca process is suppressed by a factor $e^{-\Delta/T}$ and the
modified Urca and Bremsstrahlung by a factor $e^{-2\Delta/T}$ for 
$T \leq T_c$, where $\Delta$ is the gap parameter for the CFL phase and
$T_c$ is the critical temperature below which strange matter undergoes a
phase transition into CFL matter. In this work we have used
$T_c = 0.4\Delta = 10$ MeV.

The heat produced by vortex expulsion occurs in  
an emission region just above the star's surface.  
In this region the energy, released by the magnetic field decaying
after it has been expelled from the star's interior, is deposited.
An emissivity for this process can be calculated from the
energy per unit time (cf. Eq. \ref{eq:lx}) per unit volume, 
where the volume of interest is a shell surrounding the star.
The width of the shell is estimated to be the minimum length of a
vortex still inside the star, just before it is finally expelled.
Such a vortex would be a distance of approximately
the inter-vortex spacing away from the surface of the star.
The shell width is then
\begin{equation}
\Delta R = 8.42\times 10^{-3} {\rm km}
\left(\frac{P}{1 {\rm s}}\right)^{1/4}\left(\frac{R}{10 {\rm km}}\right)^{1/2}
\left(\frac{300 {\rm MeV}}{\mu/3}\right)^{1/4} \ ,
\end{equation}
where $\mu$ is the average chemical potential throughout the star.

\begin{figure}[!t]
\begin{center}
\epsfig{file=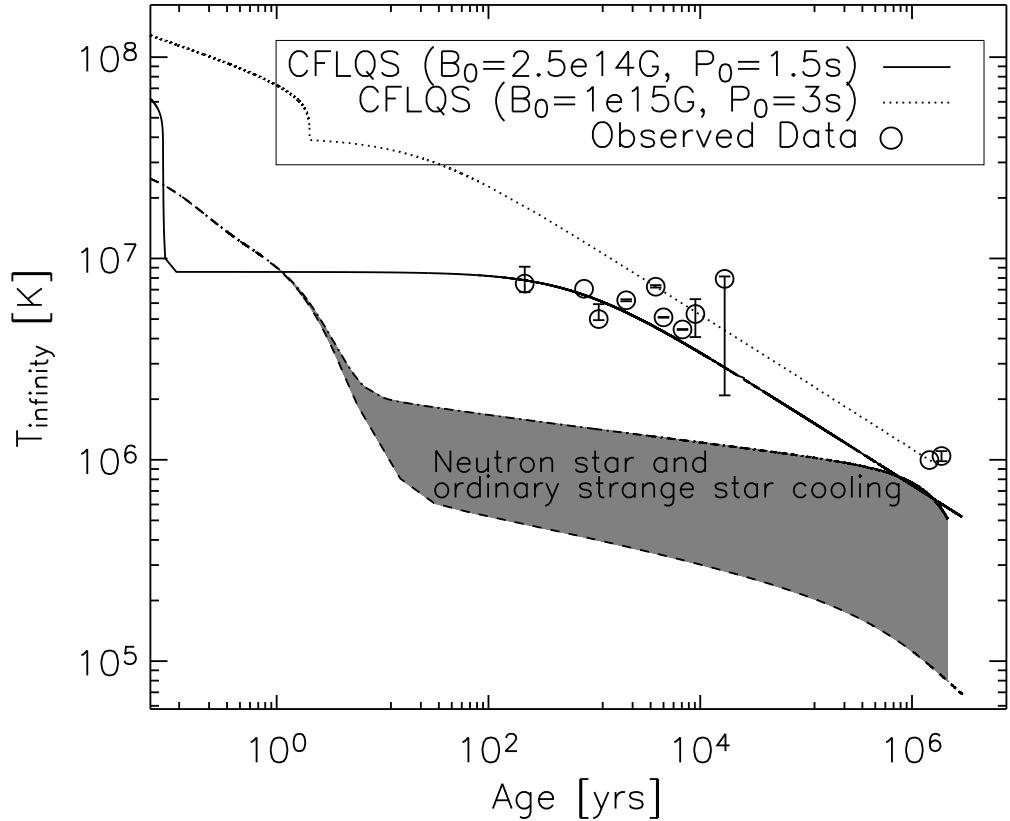,width=\textwidth}
\caption{\label{TsB15}Redshifted surface temperature evolution for
CFLQSs (CFL stars with superconductivity and the resulting vortex expulsion).
The initial spin-periods and magnetic field strengths were chosen 
such that  the evolution, due to vortex expulsion,
of spin-period, spin-down rate, magnetic field strength, and luminosity
would be consistent with observations.
The shaded region indicates calculations of cooling scenarios for various types
of neutron stars, \textit{uds} stars (strange quark stars without pairing), 
and CFL stars (without vortex expulsion).  All calculations were done assuming 
stellar masses of $1.4~M_{\odot}$ and radii of $10.5~{\rm km}$. 
The observed data is listed in Table \ref{tab:data}.}
\end{center}
\end{figure}

We have solved equations (\ref{coeq1}) and
(\ref{coeq2}) numerically for models whose structure
(composition and bulk properties) were computed from the
Tolman-Oppenheimer-Volkoff equations \cite{1996csnp.book.....G,1999Weber..book}.
The results are shown in Fig.\ \ref{TsB15}, where the redshifted 
surface temperature evolutions for various types of cooling scenarios are plotted.
These scenarios are CFLQSs (CFL stars with superconductivity and the
resulting vortex expulsion), and in the shaded region in Fig.\ \ref{TsB15},
CFL stars (without vortex expulsion), 
\textit{uds} stars (strange quark stars without pairing),
and neutron stars \cite{1999Weber..book}. The initial, or CFLQS birth, 
spin-periods and magnetic field strengths were chosen such that
the currently observed SGR/AXP/XDIN values of spin-period, 
spin-down rate, magnetic field strength, and luminosity 
would be consistent with the values derived from vortex expulsion 
(cf.\ Eq.\ \ref{eq:pdot}).  The observed data is taken from 
Table \ref{tab:data}, where ages and temperatures of SGRs, AXPs, and XDINs are listed.

Figure \ref{TsB15} shows that CFL quark stars without
vortex expulsion cool down too rapidly to agree with observed
SGR/AXP/XDIN data.  One might expect that the
suppression of emission processes, due to CFL pairing, would keep 
these stars hotter for longer, but pairing also changes the specific
heat capacity, resulting in enhanced cooling.  Standard neutron star
cooling processes also lead to stars with temperatures much lower
(albeit warmer than \textit{uds} and CFL stars without vortex
expulsion) than values observed for SGRs/AXPs/XDINs.

Emission due to vortex expulsion dominates all other
processes except during the first few minutes.  However, we also
considered the possibility that a neutron star undergoes a phase
transition to a CFLQS after a delay.  
Observational motivations for considering a delay are; (i) superluminous
 supernovae and hypernovae \cite{2008MNRAS.387.1193L}; (ii) 
the large discrepancy between SGR/AXP ages are their progenitor 
supernova remnants \cite{2009ApJ...696..562O}.
Physically, a delay may be
necessary if one considers the following; i) the time needed for a
newly born compact object to spin-down sufficiently such that the
center reaches nuclear densities \cite{2006ApJ...645L.145S}, ii)
the time for the temperature to reach the superconducting critical
value ($T_c$) and the resulting vortex lattice to form, iii) the
nucleation time for strangelets to start fusing together \cite{2007A&A...462.1017B}. 

\begin{figure}[!t]
\begin{center}
\epsfig{file=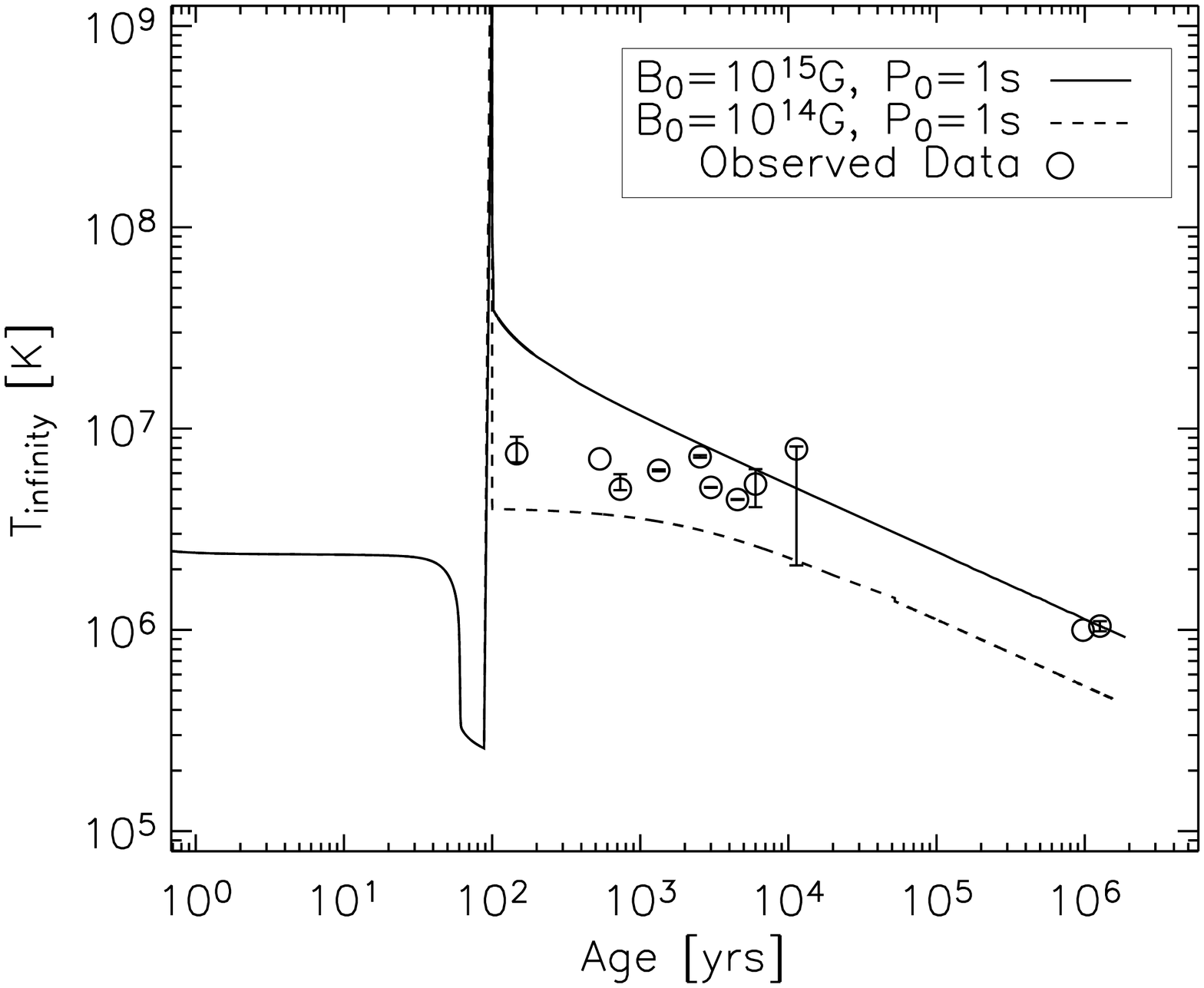,width=\textwidth}
\caption{\label{NSQS1}Thermal evolution of a neutron star undergoing a
phase transition into a CFLQS  after a delay. 
This delay is estimated to be the time needed for the central density of 
a neutron star to reach the critical value at which CFL matter is favoured ($\sim 100$ years).
Values of the birth temperature for the resultant CFLQS 
are of order $10^{11} K$, and were estimated using
the quark-nova model \cite{2002A&A...390L..39O}. The CFLQS birth periods 
and magnetic fields are as indicated.}
\end{center}
\end{figure}

Using the quark-nova model \cite{2002A&A...390L..39O} the
star is reheated to roughly $10^{11}$ K following the transition to CFL matter.
This energy is from the release of gravitational energy during the collapse
as well as the latent heat released when converting 
from hadronic to stable strange quark matter 
\cite{2002A&A...390L..39O, 2004NuPhA.735..543V, 2005ApJ...618..485K}. 
Within the framework of the quark-nova scenario the neutron stars with 
typical values for spin-period and magnetic field,
and masses greater than $1.5$ solar masses, are most likely to make the transition
to quark deconfinement and the subsequent color-superconducting phase.
Upon this transition the magnetic field may be amplified via 
color ferromagnetism to roughly $10^{15}$ G \cite{2005PhRvD..72k4003I}.  
We choose initial values for the CFLQS's magnetic field strength accordingly. 
The results are shown in Fig.\ \ref{NSQS1} for CFLQSs born
after a delay of $\tau_{\rm QN} = 100$ years.

We have computed the thermal evolution for various types of strange
quark stars and found that CFL quark stars, possessing a
rotational-vortex lattice, are in good agreement with observed
temperatures of SGRs, AXPs, and XDINs.  In our model, the CFL star
spins-down as a result of magnetic braking and expels vortices, which
contain magnetic flux, thus decreasing the magnetic field strength in
the magnetosphere, resulting in a lower spin-down rate.  Hence, the
magnetic field strength and spin period become coupled.  By including
emission from vortex expulsion in our relativistic cooling
calculations, we have found that the unusually high temperatures of
SGRs, AXPs, and XDINs can be predicted.

From previous papers \cite{2006ApJ...646L..17N, 2007A&A...476L...5N} 
we have also shown that by using our model
for a CFL quark star the evolution of the spin-period, spin-down
rate, and magnetic field strength can also be predicted for SGRs,
AXPs, and XDINs.  The long-term evolution of these properties suggests
an ancestral linkage between SGRs, AXPs, and XDINs.  We also note a
paper by \cite{2007arXiv0710.2114L}, which confirms that the birth
statistics of SGRs and AXPs are consistent with the number of observed
XDINs. Finally, this study suggests that a delay time between
supernova and quark-nova events ($\tau_{\rm QN}$) is possible.  
If the estimated delay of $100$ years is correct, 
then the density at which quark matter deconfines
can be calculated to be roughly five times the nuclear saturation density 
\cite{2006ApJ...645L.145S}.  The quark-nova delay can be measured
by considering the discrepancy between SGR/AXP ages are their progenitor 
supernova remnants.  Provided one had accurate measurements of this age 
difference, a more precise value for the density at which quark matter deconfines
could be inferred.

\bigskip
This research is supported by grants from the Natural Science and
  Engineering Research Council of Canada (NSERC), as well as the
  National Science Foundation under Grant PHY-0457329, and the
  Research Corporation. R.\ Negreiros thanks Rachid Ouyed and Brian
  Niebergal for their support and for hosting him at the University of
  Calgary during the completion of this work.

\bibliographystyle{plain}
\bibliography{ve}

\end{document}